\documentclass[11pt]{article}
\usepackage{graphicx}
\usepackage{amssymb,amsmath,epsfig}
\include{epsf}
\usepackage{amsthm}
\usepackage{color}
\usepackage{etoolbox}
\patchcmd{\footnotemark}{\stepcounter{footnote}}{\refstepcounter{footnote}}{}{}
\usepackage[colorlinks=true,urlcolor=red,citecolor=blue,hyperfootnotes=false]{hyperref}

\newcommand{\be}{\begin{equation}}
\newcommand{\ee}{\end{equation}}
\def\beqa{\begin{eqnarray}}
\def\eeqa{\end{eqnarray}}
\def\bean{\begin{eqnarray*}}
\def\eean{\end{eqnarray*}}
\def\nn{\nonumber}

\newcommand{\R}{\mathbb{R}}

\newcommand{\eqn}[1]{(\ref{#1})}
\newcommand{\del}{\partial}
\newcommand{\Tr}[1]{\:{\rm Tr}\,#1}
\newcommand{\dd}{{\mathrm d}}

\newcounter{lst}
\newenvironment{romanlist}
  {\begin{list}{\roman{lst})\ }{\usecounter{lst}
     \setlength{\labelwidth}{\leftmargin}\setlength{\labelsep}{0pt}}}
  {\end{list}}

\renewenvironment{thebibliography}[1]
         {\section*{References}\frenchspacing\small
          \begin{list}{[\arabic{enumi}]}
         {\usecounter{enumi}\parsep=2pt\topsep 0pt
         \settowidth{\labelwidth}{[#1]}
         \leftmargin=\labelwidth\advance\leftmargin\labelsep
         \rightmargin=0pt\itemsep=1pt\sloppy}}{\end{list}}

 \numberwithin{equation}{section}

\title{Metric on the space of quantum states from relative entropy.  Tomographic reconstruction }
\author{Vladimir I. Man'ko$^{a,b}$, Giuseppe Marmo$^{c,d}$, Franco Ventriglia$^{c,d}$ and Patrizia Vitale$^{c,d}$
 }

\begin{document}
\maketitle

\vspace*{-1cm}

\begin{center}

\textit{$^a$ Lebedev Physical Institute, Leninskii Prospect 53, Moscow 119991, Russia
}\\
\textit{$^b$  Moscow Institute of Physics and Technology (State University), Dolgoprudny,
Moscow Region, Russia}\\
\textit{$^c$Dipartimento di Fisica ``E. Pancini''
Universit\`a di Napoli Federico II}  \\
\textit{$^d$ INFN, Sezione di Napoli, Via Cintia 80126 Napoli, Italy}\\
 e-mail:
\texttt{manko@sci.lebedev.ru, marmo@na.infn.it, ventriglia@na.infn.it, vitale@na.infn.it}\\[1ex]

\end{center}

\begin{abstract}

In the framework of quantum information geometry, we derive, from quantum relative Tsallis entropy,   a family of quantum metrics on the space of full rank, $N$ level quantum states, by means of a suitably defined coordinate free differential calculus. The cases $N=2, N=3$ are discussed in detail  and notable limits  are analyzed. The radial limit procedure has been used to recover quantum metrics for lower rank states, such as pure states. 

By using the tomographic picture of quantum mechanics we have obtained the Fisher-Rao metric for the space of quantum tomograms and derived  a reconstruction formula of the quantum metric of density states out of the tomographic one.  
A new inequality obtained for probabilities of three spin-1/2 projections in
three perpendicular directions is proposed to be checked in experiments
with superconducting circuits.

\bigskip

\end{abstract}

\pagebreak

\section{Introduction}
In all theories where a probabilistic aspect or interpretation is present,
we face  the problem of estimating the distance between two different probability distributions.
Thus, by dealing with the abstract space of probability distributions on some sample space, we abstractly deal with all theories where probabilities play a relevant role. Probability distributions form a convex set because any convex combination of probability distributions is again a probability distribution. Those probability distributions which cannot be written as combinations are called extremal states.
From the mathematical point of view, probability distributions constitute a simplex.
The statistical-probabilistic interpretation of quantum mechanics originates from Born who provided an interpretation of the quantum mechanical wave functions as probability amplitudes.
When pure quantum states are described by rank-one projectors, it is immediate to recover that mixed quantum states are indeed a convex body whose extremal states are pure states. 
Thus not only probabilities but also probability amplitudes are a convex set. In changing the point of view from probabilities to probability amplitudes, we shift from a classical description to a quantum description, the most important aspect being now the possibility of describing interference phenomena.

It is then a natural question to look for distances on the space of probability amplitudes instead of probability distributions.
This question was indeed addressed by Wootters in his doctorate dissertation.
He found   \cite{Wootters} that a natural distance is related with the Fubini-Study metric on the space of pure states (rays of the Hilbert space associated with any quantum system).
In a relatively recent paper  \cite{Facchi} it was also shown that  the Fubini-Study metric may be related to some quantum generalization of the classical Fisher-Rao metric \cite{FR}  when wave functions are written in a polar representation and we restrict to a manifold of states characterized by a constant phase. A notion of Riemannian metrics on the space of density matrices  was discussed in \cite{dit1}.  

If in addition to the distance (or distinguishability) of two probability distributions we also consider transformations on the space of probability distributions, for instance coarse graining transformations, it is quite natural to require that the distinguishability of two distributions does not increase when two distributions are transformed into two coarser ones.This goes under the name of monotonicity requirement.
For  simplices, that is fair probability distributions, Chentsov \cite{Chentsov82} has shown that the only metric whose induced distance enjoys this monotonicity condition is the Fisher-Rao metric \cite{FR}.  In the infinite dimensional case the unicity of the Fisher-Rao metric has been proven by Michor  in \cite{michor}.
In the quantum setting, i.e.,working with probability amplitudes rather than with fair probabilities, D.Petz \cite{Petz1} has shown that metrics satisfying the monotonicity condition do not have just one representative but there are many of them characterized by an operator monotone function. 

One could say that the presence of a phase in addition to the amplitude spoils the uniqueness. Very much as for the polar representation of complex wave functions, the polar decomposition of states in a unitary matrix and a positive one, calls for the combination of a metric along the unitary orbit and a metric along the positive part. The required invariance under the unitary group allows for a relative multiplicative function between the two contributions, which has to be unitarily invariant  and depending only on the spectrum of the state which characterizes the given orbit. By factorizing an overall factor, we are left with a conformal family of metrics, each one characterized by a function \cite{LMMVV17}.
The monotonicity requirement further constraints this function so that it may be associated with a monotone operator function as discovered by Petz \cite{Petz1}.

Probabilistic theories include both Information Theory and Quantum Mechanics. It is therefore conceivable that geometry be a common playground for both, giving rise to 
Geometric Information theory, which deals   with a manifold of probability distributions (probability amplitudes for the quantum case)  on some sample space $X$, considered as a Riemannian manifold $M$ endowed with a pair of dually related connections \cite{Am85, Am00}.
\footnote{ Also see \cite{Efron} who was the first to introduce an affine connection on the parameter space manifold, and explained the role of the embedding curvature of the statistical model in the relevant space of probability distributions.}. 

It is well known that the  Hermitian metric on the space of pure states (or the Hermitian metric on the Hilbert space) may be derived from a potential, what is called  the K\"ahler potential, which gives not only the Fubini-Study metric but also the symplectic structure. It is therefore  natural to ask if the quantum metrics defined on the space of quantum states, or the space of probability amplitudes, may also be derived from a potential.  The search for a generating potential had been already addressed  by Amari \cite{Am85, Am00} in the classical framework. These potentials  often  go under different names, such as divergence functions, contrast functions or distinguishability functions.  Because of their properties, which we shall detail in Section \ref{pot}, many  relative entropies belong to the family and have been used as  potential functions.
Metric tensors and connections  are obtained  by means of a proper Taylor expansion around critical points of  potential functions.  Roughly speaking, by taking the Taylor expansion in the neighbourhood of a critical point, we have 
\be
F(x)-F(x_0)= \frac{1}{2} \frac{\del^2 F}{\del x_j \del x_k} \delta x_j  \delta x_k + \frac{1}{6} \frac{\del^3 F}{\del x_j \del x_k \del x_l} \delta x_j  \delta x_k  \delta x_l +...
\label{taylor}
\ee
Thus, the second order term may be associated with a metric, while the third order  one is related with a connection. 

As we shall see, these potential functions  are two-points functions, which  may be given very often the interpretation of cross-entropy, or relative entropy, which roughly 
speaking represent a kind of ``directed" distance between two probabilities and possess the appealing property of correlating a posteriori to a priori situations, therefore pointing at dynamics rather than statics (a property of entropy).  This procedure is well established in the classical setting but, for the quantum case, in order  to consider the Taylor expansion we would need a noncommutative differential calculus, because we should expand around states which are described by noncommutative matrices. In order to circumvent this  problem we use  exterior differential calculus on the unitary group and the quotient space of orbits, each orbit being a coadjoint orbit of the unitary group. 
In the present  paper we use  as potential function  the Tsallis relative $q$-entropy. To the best of our knowledge, the method of calculus is quite novel, and the analysis of quantum metrics associated with Tsallis entropy is also new. For particular values of the parameter $q$, known metrics are retrieved as we shall duly stress
in the paper.

We start from the consideration that a simplex of classical probabilities may be ``quantized", i.e. described in a quantum framework, by associating with every probability vector a coadjoint orbit of the unitary group acting on the dual space of its Lie algebra. This association, probability vector $\rightarrow$ density matrix, is provided by the following procedure.    A probability vector, say $(p_1, p_2, ...,p_N)$ is implemented as a density matrix by setting 
\be
\rho(U, \vec p)=U \left(
\begin{array}{ccc}
p_1  & ... & 0  \\
 0 & p_2& ... \\
 ...&...&...\\
 0&...& p_N
\end{array}
\right) U^{\dag}
\ee
In this way, we ``quantize" the classical simplex by considering the union over it of the corresponding coadjoint orbits, each one going through a probability vector identified with the diagonal elements of a density matrix. 

This union of orbits is the space of all quantum states and turns out to be a stratified manifold \cite{GKM} with the boundary being a non-smooth manifold and therefore differential geometric considerations will require additional care. We have to distinguish between orbits passing through density matrices of different ranks and those with different spectra. We shall postpone to a future paper a more thorough treatment of the subject, here we shall mainly deal with specific situations  and simple examples.

As it is well known, an alternative picture of  quantum mechanics is the quantum tomographic picture (for a recent review see \cite{introtom,pedatom} and refs. therein), which in turn can be considered from the view point of   deformation quantization \cite{MMV1, MMV2} (also see \cite{RV2012} for a derivation of an $SU(2)$ related star product).  Quantum tomograms are fair probability distributions, therefore the Fisher Rao metric can be defined  without ambiguities by means of Chentsov theorem.  In section \ref{tomo} we shall address this problem and compare the tomographic metric with the family of quantum metrics computed in section \ref{qmetric}.  Moreover, starting from the observation that there is a one to one correspondence between quantum states and tomograms through invertible maps (the so called ``quantizer-dequantizer" procedure)  we look for a reconstruction formula for quantum metrics in terms of the tomographic metric. In such a setting, unicity issues of the quantum metric are addressed. 
As an example of the tomographic picture of quantum states we present the
expression of  the density matrix of spin-1/2 system (qubit state)
in terms of three probabilities of spin-projections on three
perpendicular directions. We obtain a new inequality (uncertainty relation)
for these three probabilities and discuss a possibility of its  checking
in experiments with superconducting circuits.

The paper is organized as follows. In section \ref{pot} we review  the description of metrics and dual connections in terms of potential functions on the space of classical probabilities by introducing a suitable differential calculus.  In section \ref{qmetric} we move to the quantum setting and derive, using the relative Tsallis entropy, a new family of quantum metrics  for $N$ level systems. In section \ref{tomo} we introduce the corresponding quantum tomograms and compute the related metric. In section \ref{sympl} we switch to Gaussian symplectic tomograms, which are probability distributions defined on an infinite-dimensional statistical manifold, as opposed to the previous situation. The two appendices are dedicated to the computation of dual connections and curvature for the two-level case. We finally conclude with a short summary and perspectives.

\section{Metrics from  potentials}\label{pot}

In this section we shall review the derivation of geometric structures such as a Riemannian metric and dually related connections on the space of probability distributions \cite{Am00}. We shall also introduce a differential calculus which can be readily   extended to the quantum case. 

Let  $M$ be a space of parameters, assumed to be a smooth manifold,  characterizing a family of  states of a given system. According to \cite{Am00} and refs. therein,   a metric can be defined on such a space starting from a {\it potential function} $F$. This is   a   positive definite, convex  function  $F: M\times M \rightarrow \R$ 
\begin{romanlist}
\item
$F(x, y)> 0 \;\;\; \forall x\ne y$
\item  $F(x, y)|_{x=y } =  0 $ 
\end{romanlist}
  If we further assume $F$ to be differentiable , then $dF(x,y)|_{x=y}= 0$. 

On  considering  the diagonal immersion $i: M \rightarrow M\times M$,   condition  $ii)$ may be expressed as   $i^{*} F= 0$, so that, assuming differentiability, $ i^{*} d F= 0$. 

Being $F$ a function on $M\times M$ we shall introduce a differential calculus defined on the product manifold in terms of bi-forms\footnote{See for example \cite{doubleforms}  pag. 440 def. 9.1}. A bi-form $\gamma$ of degree $(p,q)$ is  an element of the $ {\mathcal F}(M\times M)$-module $ \Omega^p (M)\otimes \Omega^q(M)$  with $p,q\ne0$, locally given by
\be
\gamma (x,y)= \sum_{j_1..j_p; k_1,...,k_q} f(x,y)_{j_1..j_p; k_1,...,k_q} dx^{j_1}\wedge...\wedge dx^{j_p} \otimes dy^{k_1}\wedge...\wedge dy^{k_q}
\ee
and  exterior derivative 
\beqa 
&& d \otimes  {\mathbf 1} : \gamma \in \Omega^p(M)\otimes \Omega^q(M)\rightarrow  \Omega^{p+1}(M)\otimes \Omega^{q}(M) \\
&& {\mathbf 1} \otimes \tilde d : \gamma \in \Omega^p(M)\otimes \Omega^q(M)\rightarrow  \Omega^{p}(M)\otimes \Omega^{q+1}(M)
\eeqa
and we shall omit the tensor product when there is no ambiguity. This differential calculus allows for a coordinate free definition of the  metric tensor introduced in Eq. \eqn{taylor} as
\be
g(X,Y) :=  - i^*\Bigl( (d\,   \tilde d \, F)(X_l,Y_r) \Bigr)= - i^* (L_{X_l} L_{Y_r} F) \label{metricdef}
\ee
where $L_X$ indicates the Lie derivative. Specifically   the vector fields $X, Y \in \mathfrak{X}(M)$ on the left hand side (LHS)  are  identified with their image into $\mathfrak{X}(M\times M)$  on the RHS, according to  the immersion of the $\mathcal{F}(M)$-module 
$\mathfrak{X}(M)$ into the $\mathcal{F}(M\times M)$-module $\mathfrak{X}(M\times M)$ as 
$ X\rightarrow X_l \oplus \{0\}$ and $Y\rightarrow \{0\} \oplus Y_r$. Specifically, by using   the projections
\be
\begin{array}{ccc}
M\times M&\stackrel{\pi_r}{\rightarrow} &M\\
\downarrow \pi_l& &\\
M\; \; \; \;& &
\end{array}
\ee
we have  the following relations
\be
{i^l}_*(X)= X_l\;\; {\rm such\,that\;}\; L_{X_l} \pi_r^* f=0\; \;\,\; L_{X_l} \pi_l^* f=  \pi_l^*(L_X f) \;\,\; f\in\mathcal{F}(M) \;\; 
\ee
and analogous relations for $X_r$. 
We find  that $g$ is symmetric 
\be
g(X,Y)-g (Y,X)= - i^*(L_{[X_l,Y_r]} F)= dF [X_l,Y_r]_{diag}=0
\ee
and $f$-linear
\be
g(fX, Y)=- i^*(\pi^*_l f L_{X_l} L_{Y_r} F)= f \,g(X,Y) ;\; \; \;  g(X, hY)=- i^*(L_{X_l} ( \pi^*_r h  )L_{Y_r} F)= h\, g(X,Y) .
\ee
Positivity follows from $i)$ and $ii)$ because $x=y$ is a minimum. 
In local coordinates $(x,y)$ on $M\times M$ we have from \eqn{metricdef}
\be
g_{jk}= -\frac{\del^2 F}{\del x^j \del y^k} \vert_{x=y}= -i^*\left(\frac{\del^2 F}{\del x^j \del y^k}\right).
\ee
An affine connection $\nabla$ and its dual are defined on $M$ through the relations
\beqa
g(\nabla_X Y, Z)&:=& -i^* (L_{X_l} L_{Y_l} L_{Z_r} F)\\
g(\nabla^*_X Y, Z)&:=& -i^* (L_{Z_l} L_{X_r} L_{Y_r} F)
\eeqa
and the so called skewness  tensor\footnote{Let us verify that this is indeed a tensor. For that, it has to be $f$-linear in all three vector fields. We have
\beqa
T(fX,Y,Z)&=& f T(X,Y,Z)  + i^* \left[Z_l(\pi^*_r (f)) L_{X_r} L_{Y_r} F\right] = f T(X,Y,Z)  \label{eq1}\\
T(X, Y, hZ)&=& h T(X,Y,Z) -i^* \left[L_{X_l} (Y_l(\pi^*_r h) L_{Z_r}) F \right ] = h T(X,Y,Z) \label{eq2}\\
T(X, u  Y, Z) &=&  u  T(X,Y, Z) -i^* \left[\left( X_l(\pi^*_l u) L_{Y_l } L_{Z_r} -    X_r(\pi^*_r u) L_{Z_l } L_{Y_r} \right)F\right] = u T(X,Y,Z)  \label{eq3}
\eeqa
with $f,h,u\in  \mathcal{F}(M) $. The unwanted contribution in each of the equations above being zero either because the derived function does not depend on the deriving field (Eqs. \eqn{eq1} and \eqn{eq2}), or (Eq. \eqn{eq3})  because it is the difference of two terms which coincide when evaluated at $x=y$. } \cite{Am00}
\be
T(X,Y,Z)= g(\nabla_XY-\nabla_X^* Y, Z), \;\; X,Y,Z \in\mathfrak{X}(M). \label{skewness}
\ee
which is symmetric.  In local coordinates we have
\beqa
\Gamma_{ljk}(x)&=& -\frac{\del^{3} F}{\del x_j \del x_k \del y_l} \vert_{x=y} \nn\\
\Gamma^*_{ljk}(x)&=& -\frac{\del^{3} F}{\del y_j \del y_k \del x_l} \vert_{x=y} \nn\\
T_{ljk} (x)&=& \Gamma_{ljk}(x)- \Gamma^*_{ljk}(x)
\eeqa
The two connections $\nabla, \nabla^*$ are easily checked to be torsionless. Indeed we have
\be
0= g(\nabla_XY-\nabla_YX, Z)= -i^*(L_{Z_r}(L_{X_l} L_{Y_l}- L_{Y_l} L_{X_l}) F)
\ee
and analogously for $\nabla^*$. 

Manifolds of interest in information geometry are \cite{Am00}  {\it statistical manifolds} $P(\chi)$ of probability densities $p(x)$  on a measure space $\chi, x\in \chi$, and, in particular,  the spaces of parameters of  {\it statistical models}. The latter are families of probability densities $p(x;\xi)$ parametrized by a set of n variables $\xi=[\xi_1, ..., \xi_n]$, so that the map $\xi \rightarrow p(x;\xi)$ is injective. The  space of parameters is required to carry the structure of a differential manifold and can be equipped with the two tensors defined above, the metric $g$ and the skewness tensor $T$. The triple $(\Sigma, g, T)$ is usually referred to as a  statistical model.

As an example, let us consider $p\in \mathcal{P},   \{p=p_1,...,p_n\}$ with $p_i\equiv p(x_i), i=1,...,n $ the probability distribution associated to the value $x_i$ of a discrete random variable $X$. We can define a potential by means of the Shannon entropy $H=- \sum_j p_j \log p_j$, and introducing the {\it relative entropy}
\be
H(p, q)= \sum_j p_j (\log p_j- \log q_j)\;\;\, p,q\in \mathcal{P}. \label{relShan}
\ee
As an alternative, we could also consider the potential function
\be
F(p,q) = 4(1-\sum_j \sqrt{p_j q_j}) \label{altpot}
\ee
It is possible to verify that both these  potentials define the same metric on the parameter space, which in this case is identified with  the statistical manifold itself, $\mathcal{P}$. 
Indeed we have from Eq. \eqn{relShan}
\be
g^H_{jk}(p) = -\frac{\del^2 H (p,q)}{\del  p_j \del q_k }\vert_{p=q} = \delta_{jk} \frac{1}{p_k}
\ee
while from Eq. \eqn{altpot} 
\be
g^F_{jk}(p) = -\frac{\del^2F (p,q)}{\del p_j \del q_k }\vert_{p=q} = \delta_{jk} \frac{1}{\sqrt{p_k q_k}}\vert_{p=q}= \delta_{jk} \frac{1}{p_k}
\ee
both yielding  the Fisher-Rao metric
\be
g= \sum_j p_j d \ln p_j \otimes d\ln p_j. \label{FR}
\ee
On the contrary we obtain for the connection
\beqa
\Gamma^H_{jkl}(x)&=& (\Gamma^H)^*_{jkl}(x)=0 \nn\\
\Gamma^F_{jkl}(x)&=& (\Gamma^F)^*_{jkl}(x)=-\frac{1}{2 p_k^2}\delta_{jl} \delta_{jk}
\eeqa
which is flat in both cases.  The divergence function  in Eq. \eqn{altpot} possesses some interesting properties. We shall come back to it in Section \ref{uncert}. 

\section{Quantum metric  from the relative Tsallis entropy}\label{qmetric} 

Having in mind the geometric description of Quantum Mechanics, the ``potential function'' description turns out to be very convenient also to describe metrics on the space of quantum states. The states of a given quantum system are  characterized in terms of density operators. The parameter space $M$ is now given by the space of parameters, $\{\xi\}$, identifying a family of  states $\rho(x, \xi) $. Analogously to  the classical situation analyzed in the previous section we can use  relative entropy functions as the starting potential. 

To this end let $\mathfrak{S}(\rho,\tilde{\rho})$ be the relative Tsallis entropy
\be
\mathfrak{S}(\rho,\tilde{\rho})= (1-\Tr\rho^q\tilde{\rho}^{1-q})(1-q)^{-1} \label{Tsal}
\ee
with $q$ a positive real parameter and $\rho, \tilde \rho$ representing two different quantum states. This is nothing but an $\alpha$ divergence function with $\alpha = 2q-1$ (see \cite{Am00} for details).  As such, it is closely related to the    R\'enyi relative entropy  (see  \cite{takahashi} for recent applications). 

  $\mathfrak{S}$ is a function in $\mathcal{F}(M\times M)$, with $M$ the parameter space.   The Tsallis entropy can be regarded as  a regularization of the von Neumann entropy, with the logarithm replaced by the q-logarithm function
\be
\log_q \rho= \frac{1}{1-q}(\rho^{1-q}-1).
\ee
Indeed, in the limit $q\rightarrow 1$ Eq. \eqn{Tsal}  reproduces the relative von Neumann entropy:
\be
\lim_{q\rightarrow 1} (1-\Tr\rho^q\tilde{\rho}^{1-q})(1-q)^{-1}= \Tr\rho(\ln \rho-\ln\tilde{\rho}).\label{vonN}
\ee
This function $\mathfrak{S}$  satisfies conditions stated at the beginning of section \ref{pot}, that is, it characterizes as a potential function.  
Inspired by Petz \cite{Petz1, Andai2003}, let us  introduce the   divergence function 
\be
{S}_{Ts}(\rho,\tilde{\rho})= \frac{1}{q} \mathfrak{S}_{Ts}(\rho,\tilde{\rho}) \label{symS}
\ee
which is equal to the Tsallis relative entropy rescaled by a factor of $1/q$.  This function is symmetric under the exchange $\rho\rightarrow \tilde{\rho}$, and $q\rightarrow (1-q)$. Moreover, 
 it gives  the same $q\rightarrow 1$ limit as $\mathfrak{S}_{Ts}$, but, for $q\rightarrow 0$, the expression
\be
\lim_{q\rightarrow 0 } q^{-1} \mathfrak{S}_{Ts}(\rho,\tilde{\rho})= \Tr\tilde \rho(\ln \tilde \rho-\ln\rho)
\ee
which is  the von Neumann relative entropy with respect to $\tilde{\rho}$.  Therefore, we choose to use the rescaled function $S_{Ts}(\rho,\tilde{\rho})$ for the rest of the article. 

From now on, we shall consider $N$-level (finite) quantum systems. The space of parameters $M$ coincides with the space of states itself. This is a stratified manifold,  with strata determined by the rank of the states and  given by   the union of unitary orbits of $SU(N)$, of different dimension \cite{GKM, Sc03}. We shall study in detail the cases $N=2$ and $N=3$. Unitary
orbits can be naturally identified with spheres in $R^{N^2-1} $ only for $N=2$, where the total space is therefore the Bloch  ball,  while in higher dimensions the embedding of quantum states into  a closed ball in $R^{N^2-1} $ is not surjective. Unitary orbits  have different dimensions and they are not in one-to-one correspondence with spheres.  Therefore the case $N=3$  is particularly interesting as an instance of such a fundamental difference. 

The density matrices $\rho, \tilde \rho$ can be parametrized  in terms of diagonal matrices and  unitary transformations
\be
\rho= U\rho_0 U^{-1}, \;\;\; \tilde \rho= V\tilde{\rho}_0 V^{-1} \label{dmatrix}
\ee
where $ U,V\in SU(N)$ are special unitary $N\times N$ matrices, with $N$ labeling the levels.  Clearly, the unitary matrices $U,V$  are determined only up to  unitary transformations by elements in the commutant of $\rho_0$ or $\tilde{\rho}_0$.The diagonal matrices associated with states, form a simplex, therefore, if we limit the analysis to faithful states, we can consider the space to be parametrized by some homogeneous space of $SU(N)$, i.e., $SU(N)$ quotiented by the stability group of the state times the open ``part" of the simplex to which the diagonal part of the state belongs. As the homogeneous spaces of $SU(N)$ are not parallelizable , we shall consider the differential calculus we are going to use as carried on the group $SU(N)$ times the open part of the simplex. We shall apply it to the subalgebra of covariant tensor fields generated by the functions $f$(states) pulled-back to  $SU(N)$ times the interior of the simplex. 

According to the definition given in eq. \eqn{metricdef} we have
\be
g=-i^*d\, \tilde d S_{Ts}(\rho,\tilde{\rho})=\left(q(1-q)\right)^{-1}  i^{*}\Tr d\rho^q\otimes \tilde d\tilde \rho^{1-q} \label{genmet}
\ee
with 
 exterior derivatives  respectively acting on $\rho, \tilde{\rho}$. When acting on functions defined on $M\times M$ they can be defined as follows
 \be d= \theta^j  L_{X_j}, \;\;\;\;\; \tilde d= \tilde\theta^j  L_{\tilde X_j} \label{diffs}
\ee
 with  $(X_j, \theta^j)$, $(\tilde X_j, \tilde \theta^j)$, dual  bases of vector fields and one-forms on each copy of $M$\footnote{Notice that when acting on generic p-forms,  \eqn{diffs} have to be generalized to $d= \theta^j \wedge L_{X_j}- \theta^j\wedge di_{X_j}$ and similarly for $\tilde d$. Indeed, from the definition $L_X= i_X d + d i _X$, we have $\theta\wedge L_X=\theta\wedge i_X d + \theta\wedge d i _X$ from which the result, observing that, on a basis $ \theta^j\wedge i_{X_j} =1$  }. The tensor product is to be understood in the space of one-forms, providing us with a bi-form, whereas the trace is defined on the space of  $N\times N$ matrices. 
We now use
\beqa
d \rho^q&=& d(U\rho_0^q U^{-1})=dU\rho_0^q U^{-1}+ U d\rho_0^q U^{-1}- U\rho_0^q U^{-1}d U U^{-1} \\
\tilde d \tilde{\rho}^{1-q}&=& d(V\tilde{\rho}_0^{1-q} V^{-1})=dV\tilde{\rho}_0^{1-q} V^{-1}+ V d\tilde{\rho}_0^{1-q} V^{-1}- V\tilde{\rho}_0^{1-q} V^{-1}d V V^{-1}
\eeqa
and perform the pullback to the manifold M (which amounts to put $\tilde{\rho}_0=\rho_0$ and $U=V, U^{-1}=V^{-1}$), 
so to have
\be
g_q = \left(q(1-q)\right)^{-1} \Tr \Bigl( [U^{-1} dU, \rho_0^q]\otimes   [U^{-1} dU, \rho_0^{1-q}]+ q(1-q) \rho_0^{-1} d\rho_0\otimes d\rho_0 \Bigr) \equiv g_q^{tan}+ g_q^{trans}\label{generalg}
\ee
which, due to the lack of mixed terms, may be split  into ``tangent''  and ``transversal''  part to the orbit of the unitary group.  Eq. \eqn{generalg} is one of the main findings of the present paper. 
It  is a general result for $N$ level systems.  The approach we have followed in order to obtain the family of metrics in Eq. \eqn{generalg} is not restricted to Tsallis relative entropies but can be generalized to any other potential function. We shall come back to the subject in a forthcoming paper.

In the next section we shall specialize to definite examples with  $N=2,3$. We notice in eq. \eqn{generalg} that we have two distinct contributions, the first one is the tangent contribution  to the orbits of $SU(N)$ in the space of states, at fixed $\rho_0$; this is a purely quantum term, which is related to the ``phase''; the second one instead represents the metric in the direction transversal to the orbits; this is a ``classical'' part.  Let us notice that the transversal part is certainly degenerate for pure states. 

Let us recall that, for any Lie group n-dimensional Lie group $G$, realized as group of matrices, the fundamental Maurer-Cartan one-form $g^{-1}dg := \theta_j \tau^j$ yields   a basis of left-invariant one forms $\theta_j, j= 1, ...,n$, while $\tau^j, j=1,...,n$  are a basis for the matrix Lie algebra of $G$.  When $G$ is the unitary group $\theta_j$ are the analogue of phases for the wave functions expressed in polar coordinates. 

\subsection{The limit $q\rightarrow 1$}
in order to recover  the metric obtained from the relative von Neumann entropy \eqn{vonN}, 
  it is instructive  to perform the limit $q\rightarrow 1$ in Eq. \eqn{generalg}.  The transversal part, related to diagonal matrices,  simply gives
\be
g_1^{trans}= \Tr \rho_0^{-1} d\rho_0\otimes d\rho_0= \Tr \rho_0 d \ln\rho_0\otimes d\ln\rho_0
\ee
whereas the tangential part  is given by 
\beqa 
g_1^{tan}&= & \lim_{q\rightarrow 1} \left(q(1-q)\right)^{-1}\Tr \Bigl(\rho_0^{1-q} U^{-1} dU \rho_0^q\otimes   U^{-1} dU- U^{-1} dU \rho_0 \otimes U^{-1} dU \Bigr) \nn\\
&+&  \lim_{q\rightarrow 1} \left(q(1-q)\right)^{-1}\Tr \Bigl( U^{-1} dU\otimes  \rho_0^{1-q}U^{-1} dU\Bigr)\rho_0^q-   U^{-1} dU \otimes   U^{-1} dU \rho_0 \Bigr)\nn\\
&= &  \lim_{q\rightarrow 1}\Tr  \Bigl( \ln\rho_0 \rho_0^{1-q} U^{-1} dU\otimes  \rho_0^{q}U^{-1} dU -\rho_0^{1-q} U^{-1} dU
\ln\rho_0 \rho_0^{q}\otimes  U^{-1} dU \Bigr. \nn\\
&+&\Bigl.  U^{-1} dU \ln\rho_0 \rho_0^{1-q} \otimes U^{-1} dU \rho^q_0 - U^{-1} dU\rho_0^{1-q}\otimes U^{-1} dU \rho_0^q  \ln\rho_0  \Bigr) 
\eeqa
Performing the limit we finally obtain 
\be
g_1= \Tr \rho_0^{-1} d\rho_0\otimes d\rho_0+  \Tr [U^{-1} dU, \ln \rho_0] \otimes [U^{-1} dU,  \rho_0] .\label{q=1met}
\ee
We shall come back to this expression later on. 

Let us  notice that,  in the limit $q\rightarrow 0$ we get exactly the same result. 

The  general treatment illustrated in this section is closer in spirit to what is known in the literature as ``nonparametric'' description \cite{Pistone}. In what follows we shall consider states parametrized by real differential manifolds \cite{GKM2}
so that they carry a differential calculus. 

\subsection{Two level systems}
For a two level system the relevant group of unitary transformations is $SU(2)$. The  matrix valued one-form $U^{-1} dU$ is the left invariant one-form on the group and can be written in terms of Pauli matrices and basic left-invariant one forms:
\be U^{-1} dU= i \sigma_j \theta^j. \label{mcartan}\ee
The diagonal density matrix $\rho_0$ is characterized in terms of a single real parameter $-1 \le w\le 1$,  that is  
\be
\rho_0= \frac{1}{2}(\sigma_0+ w \sigma_3).
 \ee
 with $\sigma_0$ the identity matrix. 
We have then 
\be
\rho_0^q=\left(
\begin{array}{cc}
(\frac{1+w}{2})^q  &  0  \\
 0 &   (\frac{1-w}{2})^q 
\end{array}
\right) 
\ee
and similarly for $\rho_0^{1-q}$, that is
\beqa 
\rho_0^q &=&  \frac{1}{2}(a_q+b_q)  \sigma_0+  \frac{1}{2}(a_q-b_q) \sigma_3 \label{rho0q}
 \\
\rho_0^{1-q} &=&  \frac{1}{2} (a_{1-q}+b_{1-q} ) \sigma_0+\frac{1}{2} (a_{1-q}- b_{1-q}) \sigma_3 \label{rho01-q}
\eeqa
with  
\beqa
a_q&=& (\frac{1+w}{2})^q \nn\\
b_q &=& (\frac{1-w}{2})^q 
\eeqa
and analogous expressions for $a_{1-q}, \, b_{1-q}$.  In the two dimensional case  the full density matrix \eqn{dmatrix} can also be written  as 
\be \rho= U\rho_0 U^{-1} = \frac{1}{2} (\sigma_0 +w\, \vec x\cdot \vec \sigma) \label{rho2}
\ee
with the parameters $x_i$ functions of  the unitary matrix elements through the relation 
\be
U \sigma_3 U^{-1}= \vec x \cdot \vec \sigma   \label{xes}
\ee
which implies, on taking the square on both sides, $\sum_i x_i x_i = 1$.  The manifold of parameters is therefore the three dimensional ball $B^2$ with radius equal to 1. This is indeed a stratified manifold, with two strata, one  given by pure states , $w^2=1$, the other by  the union of unitary orbits of $SU(2)$, with $0\le w^2<1$,  passing through rank two states  \cite{ GKM, Sc03}.
On replacing \eqn{rho0q}, together with \eqn{mcartan} in the expression for the quantum metric \eqn{generalg} we finally arrive at
\be
g_q= \left(q(1-q)\right)^{-1}\Bigl(q (1-q) \frac{1}{1-w^2} dw\otimes dw +2 (a_q-b_q)(a_{1-q} - b_{1-q})  (\theta^1\otimes\theta^1+\theta^2\otimes\theta^2) \Bigr) \label{n=2metric}
\ee
where we have used, for the contribution which is tangent to $SU(2)$ orbits, the relation
\be
[U^{-1}dU, \rho_0^q]= (b_q-a_q)   \theta^j \epsilon_{j3}^k \sigma_k
\ee
with the analogous one for $\rho_0^{1-q}$, and $\sigma_j \sigma_k= \delta_{jk}+i \epsilon_{jk}^l\sigma_l$. 
Let us recall  that, when  $\rho$, $\tilde{\rho}$ commute, they can be simultaneously diagonalized, yielding just the  transversal contribution, which coincides with the Fisher Rao metric. In fact, posing $\frac{1+ w}{2} = k_1, \; \frac{1- w}{2} = k_2$ we have
\be
 \sum_j k_j d\ln k_j\otimes d\ln k_j= \frac{1}{k_1} d k_1\otimes d k_1 + \frac{1}{k_2} d k_2\otimes d k_2 = \frac{1}{1-w^2} dw\otimes dw
 \ee
which coincides with  the first term of Eq. \eqn{n=2metric}.  This result confirms the utility of introducing the rescaling of $1/q$ in the definition of the Tsallis entropy.  In terms of  the parametrization $k_1, k_2$ the metric \eqn{n=2metric} is finally  rewritten  as
\be
g_q=  \sum_j k_j d\ln k_j\otimes d\ln k_j + 2 \frac{k_1^q-k_2^q}{q} \frac{k_1^{1-q}-k_2^{1-q}}{1-q} (\theta^1\otimes\theta^1+\theta^2\otimes\theta^2)   \label{k1k2met}
\ee
which will be used for higher dimensional comparison.

For two level systems we only have two strata. One is provided by rank two-states, the other one by pure states, characterized by $w^2=1$.  In the chosen parametrization  Eq. \eqn{n=2metric} only holds inside the Bloch ball, that is, for invertible density states. In order to recover the metric $g_q^0$ for pure states   living on the boundary of the Bloch ball, we follow the procedure described in \cite{Petz-Sudar, Sudar} where a weak radial limit is performed. This amounts to evaluating the scalar product of two tangential vectors, say $\hat A, \hat B$, at a point $\rho_D$ strictly  inside the Bloch sphere,  yielding $g_{q}(\hat A, \hat B)\vert_{\rho_D}$;  thus, only the tangential part of the metric contributes.  To be definite let us choose  the density state $\rho_D$  with $k_1 > k_2$.   Then, we perform the radial limit along the radius passing through $\rho_D$, up to the pure state $\rho_P$ with eigenvalue $k_1=1, k_2=0$. This limiting procedure yields
 \be
 g_q^0= \left(q(1-q)\right)^{-1} (\theta^1\otimes \theta^1 + \theta^2\otimes \theta^2) \label{radlim}
 \ee
 which is singular for $q\rightarrow 1, 0$.  
 
For  $q=1$ let us first derive the expression for rank two density states (the case $q=0$ can be treated in the same way). This is    obtained by performing the limit in Eq. \eqn{n=2metric} (or directly by eq. \eqn{q=1met}) to be
\be
g=  \frac{1}{1-w^2} dw\otimes dw+  2 w\ln\frac{1+w}{1-w} (\theta^1\otimes\theta^1+ \theta^2\otimes\theta^2) \label{n=2q=1metric}
\ee
which coincides  with the result of \cite{Ercolessi2, Sam}. 
We recognize in Eqs. \eqn{n=2metric}, \eqn{n=2q=1metric}, the transversal contribution and the round metric of the sphere $S^2$, with different coefficients.   We observe that for $q=1,0$, in agreement with  Eq. \eqn{radlim},   it is not possible to perform the radial limit procedure  to recover the metric for pure states because the coefficient of the tangential component diverges for $w\rightarrow \pm 1$  (namely the radial limit and the limit $q\rightarrow 1,0$ commute and give a negative result). 

 The one-parameter families of metrics that we have obtained in Eqs. \eqn{n=2metric}, \eqn{n=2q=1metric} are in agreement with the Petz classification theorem \cite{Petz1} that establishes a correspondence between monotone metrics and operator monotone functions $f: [0,\infty[ \rightarrow \R$, such that $f(t)=t\, f(1/t)$ hold for all positive $t$, with,  for the two-levels case,  monotone metrics written in the following form: 
\be
g= \frac{1}{1-w^2} dw\otimes dw + \frac{w^2}{(1+w) f\Bigl(\frac{1-w}{1+w}\Bigr)} (\theta^1\otimes\theta^1+\theta^2\otimes\theta^2)
\ee
In order  to prove that, it is convenient to pose 
\be
 t=\frac{1-w}{1+w}\,.
\ee

Thus, the function $f$ reproducing Eqs.  \eqn{n=2metric}, \eqn{n=2q=1metric} is easily checked to be \cite{Petz2, Andai2003}
\be
f(t)= \frac{ q }{t^q-1} \frac{1-q}{t^{1-q}-1}  (t-1)^2
\ee
for $q\ne 1,0$ and 
\be
f(t)= \frac{t-1}{\ln t}
\ee
for $q=1,0$, up to a normalization factor of $1/4$.  Let us observe that the radial procedure described above can be  understood in terms of the function $f(t)$.  According to \cite{Petz-Sudar, Sudar} the radial limit is well defined  if and only if $f(0)\ne 0$. This is indeed the case for  $q\ne 1,0$ but it is not verified for $q=1,0$. 

Finally, for completeness and to help the comparison with similar computations in information geometry \cite{Am00, dit2, dit3, dit4,  Gib2, Gib3}, we conclude the discussion of the two-dimensional case with the calculation of dual connections and related curvatures. The results are shown in the appendices.

Since for a two level system the topological boundary coincides with the extremal case, we shall consider in the next section the case of three level systems where the boundary of the open submanifold of states with maximal rank contains two different strata of rank one and two.

\subsection{Three level systems}
For three level systems the relevant group of unitary transformations is $SU(3)$. The parameter manifold $M$ is a proper submanifold of $3\times 3$ matrices.  The strata are  the union of  unitary orbits of $SU(3)$ (four- and six-dimensional sub-manifolds in $\R^8$) passing through rank one, two and three density states. 

The left invariant matrix-valued Maurer-Cartan one-form $U^{-1} dU$ can be written in terms of Gell-Mann matrices\footnote{We choose the following basis of traceless matrices: \bean && \lambda_1=\left(\begin{array}{ccc}
0& 1&0\\1& 0&0\\ 0& 0& 0
\end{array} \right),\; \lambda_2=\left(\begin{array}{ccc}
0&-i &0\\i& 0&0\\ 0&0 &0 
\end{array} \right),\;\lambda_3=\left(\begin{array}{ccc}
1&0 &0\\0&-1 &0\\0 &0 & 0
\end{array} \right),\; \lambda_4=\left(\begin{array}{ccc}
0&0 &1\\0& 0&0\\ 1& 0&0 
\end{array} \right),\\
&&  \lambda_5=\left(\begin{array}{ccc}
0& 0&-i\\0&0 &0\\ i& 0& 0
\end{array} \right),\;  \lambda_6=\left(\begin{array}{ccc}
0&0 &0\\0& 0&1\\0 & 1& 0
\end{array} \right),\;  \lambda_7=\left(\begin{array}{ccc}
0&0 &0\\0&0 &-i\\ 0&i & 0
\end{array} \right),\;  \lambda_8=\frac{1}{\sqrt 3}\left(\begin{array}{ccc}
1&0 &0\\0&1 &0\\ 0& 0& -2
\end{array} \right)
\eean}
 $\lambda_j, j=1,...,8$:
\be U^{-1} dU= \lambda_j \theta^j \; \; j=1,...,8 \label{mcartan3}\ee
with $\theta^j$ the basic left-invariant one forms, as in the previous case.
The diagonal density matrix $\rho_0$ may be written in terms of three real  parameters  $k_1, 
k_2, k_3>0,$ with the constraint $k_1+ k_2+ k_3= 1$. We have
\be
\rho_0^q=\left(
\begin{array}{ccc}
k_1^q  &  0 &0 \\
 0 &   k_2^q&0\\
0&0& k_3^q
\end{array}
\right) 
\ee
and a similar expression  for $\rho_0^{1-q}$. The matrix $\rho_0$ can be expressed in terms of the traceless  diagonal matrices of $\mathfrak{su}(3)$,  which in the chosen basis are   $\lambda_3$ and $ \lambda_8$,  and the $3\times 3 $ identity matrix, $\lambda_0$,
\beqa
\rho_0^q &=&  (\alpha_q \lambda_0+ \beta_q \lambda_3+ \gamma_q\lambda_8) \\
\rho_0^{1-q} &=&  (\alpha_{1-q} \lambda_0+ \beta_{1-q} \lambda_3+\gamma_{1-q}\lambda_8)
\eeqa
with 
\beqa
\alpha_q&=& \frac{1}{3}\Bigl(k_1^q+k_2^q + k_3^q \Bigr)\nn\\
\beta_q &=& \frac{1}{2}(k_1^q-k_2^q)\nn\\
\gamma_q &=& \frac{k_1^q+k_2^q -2 k_3^q}{2\sqrt{3}}
\eeqa
and analogous expressions for $\alpha_{1-q}, \, \beta_{1-q},\gamma_{1-q}$. 
We have thus 
\be
\rho= U\rho_0 U^{-1} = \lambda_\mu x^\mu = (\frac{1}{3} \lambda_0 + x^j \lambda_j) \label{rho3}
\ee
with $\mu=0,...,8$ and $\lambda_0$ the identity matrix ${\mathbf 1}_3$. The condition $\Tr \rho= 1$ imposes $x^0 = 1/3$. In order to identify the manifold of parameters, we repeat the analysis of the two-levels case. From \eqn{rho3} we have
\be
U(\beta_1 \lambda_3+ \gamma_1\lambda_8)U^{-1} =x^j \lambda_j,\;\; j=1, ...,8
\ee
which implies, after taking the square on both sides and the trace
\be
2(\beta_1^2+\gamma_1^2)= 2 \sum_1^8 x_j x_j
\ee
or, in terms of the parameters $k_1, k_2, k_3$, with the constraint $k_1+k_2+k_3=1$
\be
\sum_1^8 x_j x_j=\frac{ 1}{3} -(k_1 k_2 +k_2 k_3 + k_1 k_3)
\ee
which  identifies the parameter manifold as a submanifold with boundary in $R^8$. Details on the manifold of parameters can be found in \cite{Sc03, EMMM2007}.
Let us now evaluate  the quantum metric \eqn{generalg} for this case. The transversal term is easily computed to be
\be
\Tr \rho_0^{-1} d\rho_0\otimes d\rho_0=  \sum_{j=1}^3 \frac{1}{k_j} dk_j\otimes dk_j  =  \sum_j k_j d\ln k_j\otimes d\ln k_j. \label{transverseq3}
\ee
Notice that the limit $q\rightarrow 1,0$ is regular. 

In order to compute the term tangential to the unitary  orbits in \eqn{generalg} we
use 
\be
[U^{-1}dU, \rho_0^q]= 2i  \theta^j (\beta_q f_{j3}^k\lambda_k+ \gamma_q f^k_{j8} \lambda_k) 
\ee
with $f^{lm}_n$ the structure constants of $SU(3)$\footnote{We have $[\lambda_i, \lambda_j]= i f_{ij}^k \lambda_k$, where the only non-zero structure constants, are, for our choice of the Lie algebra basis,
$ f^{123}=2, f^{147}=f^{165}=f^{246}=f^{257}=f^{345}=f^{376}=1 \,;\;\; 
f^{458}=f^{678}=\sqrt{3} $}. 
On using these expressions, together with \eqn{mcartan3}  we finally arrive at
\beqa
&& \Tr  [U^{-1} dU, \rho_0^q]\otimes   [U^{-1} dU, \rho_0^{1-q}]=  8 \beta_q \beta_{1-q} \sum_{j\ne 3,8}\theta^j\otimes\theta^j +6 (\gamma_q\gamma_{1-q}- \beta_q \beta_{1-q}) \sum_{j\ne 1,2,3,8}\theta^j\otimes\theta^j  \nn\\
&+& 2{\sqrt 3} (\beta_q\gamma_{1-q}+\beta_{1-q}\gamma_q) (\sum_{j=4,5}\theta^j\otimes\theta^j  -\sum_{j=6,7}\theta^j\otimes\theta^j ) \label{tangentq3}
\eeqa
so that the total metric is given by the sum of the two contributions 
\be
g_q= \rm{Eq.}\eqn{transverseq3}+\rm{Eq.}\eqn{tangentq3}.
\ee
The case $q=1$ is obtained as a limit to be
\beqa
g&=&  \sum_{i=1}^3 k_i d\ln k_i\otimes d\ln k_i + 8 \bar\beta \beta \sum_{j\ne 3,8}\theta^j\otimes\theta^j   \nn\\ 
&+ &  \left(6 (\gamma\bar \gamma- \beta \bar \beta)\right)\sum_{j\ne 1,2,3,8}\theta^j\otimes\theta^j  +2{\sqrt 3}  (\beta\bar\gamma+\bar\beta\gamma) (\sum_{j=4,5}\theta^j\otimes\theta^j  -\sum_{j=6,7}\theta^j\otimes\theta^j ) 
\eeqa
with 
\beqa
&& \bar\beta= \frac{1}{2}\ln\frac{k_1}{k_2}, \;\;\; \;\;\bar\gamma= \frac{\ln k_1 +\ln k_2-2 \ln k_3}{2\sqrt{3}} \nn\\
&&\beta= \frac{k_1-k_2}{2}, \;\;\;\; \gamma = \frac{ k_1+  k_2-2 k_3 }{2\sqrt{3}} .
\eeqa
Let us have a closer look at the tangential term given  in Eq. \eqn{tangentq3}.  On using the explicit expressions for $\beta$ and $\gamma$ in terms of the parameters $k_1,k_2,k_3$ we obtain
\beqa
g_{q}&=&  \sum_{i=1}^3 k_i d\ln k_i\otimes d\ln k_i \nn\\
&+&\frac{2}{q(1-q)} [\,(k_1^q-k_2^q)(k_1^{1-q}-k_2^{1-q})(\theta^1\otimes\theta^1+\theta^2\otimes\theta^2) \nn\\
&&\;\;\;\;\;\;\;\;\; \;\;\;\; + (k_1^q-k_3^q)(k_1^{1-q}-k_3^{1-q})(\theta^4\otimes\theta^4+\theta^5\otimes\theta^5) \nn\\
&&\;\;\;\;\;\;\;\;\; \;\;\;\; + (k_2^q-k_3^q)(k_2^{1-q}-k_3^{1-q})(\theta^6\otimes\theta^6+\theta^7\otimes\theta^7) ]\label{gtgorb}
\eeqa
which makes evident the splitting into three $SU(2)$ copies (cfr. Eq. \eqn{k1k2met}). 
For $k_1\ne k_2\ne k_3$ we have that the tangential part of the metric is the pullback of a six-dimensional orbit of $U(3)$, corresponding to $U(3)/U(1)\times U(1) \times U(1)$,  whereas for each two of them being coincident, but different from  the remaining one, that is $k_j=k_l\ne k_m, j\ne l \ne m=1,2,3$  we find  four-dimensional unitary orbits which correspond to $U(3)/U(2)\times U(1)$ \cite{Sc03, Ercolessi2}.  

\noindent {\bf Remarks}. We observe that the radial limit procedure illustrated in the two-dimensional case can be performed also in the present case for rank one (pure states)  and rank two density states. For pure states,  by using horizontal vectors which are tangent to the orbits,  we first get rid of the transversal component of the metric and then, choosing for example  $k_1> k_2, k_3$ we perform the limit    $k_1\rightarrow 1, k_2,k_3\rightarrow 0$. We thus obtain 
\be
g^0_q =  \frac{2}{q(1-q)} \sum_{j\ne 3,6,7,8}\theta^j\otimes\theta^j
\ee
which is singular for $q\rightarrow 1,0$ as in the two-dimensional case discussed in the previous section. Moreover, in the present situation we   can perform another weak limit by evaluating the metric over ``partially transversal'' vectors, namely those  whose transversal part is zero when evaluated over e.g.   $dk_3$. Then, for the remaining part of the metric, choosing $k_1 > k_2> k_3$ we can perform the limit $k_3\rightarrow 0, k_1+k_2\rightarrow 1$. In this way we obtain the metric for the stratum of rank two density states, which reads 
\beqa
\tilde g_q&=&  \frac{2}{q(1-q)}  [\,(k_1^q-k_2^q)(k_1^{1-q}-k_2^{1-q})(\theta^1\otimes\theta^1+\theta^2\otimes\theta^2) \nn\\
&&\;\;\;\;\;\;\;\;\; \;\;\;\; + k_1(\theta^4\otimes\theta^4+\theta^5\otimes\theta^5) \nn\\
&&\;\;\;\;\;\;\;\;\; \;\;\;\; +   k_2(\theta^6\otimes\theta^6+\theta^7\otimes\theta^7)]_{k_1+k_2=1}.
\eeqa
It is straightforward to check that if we further perform the radial weak limit  $k_2\rightarrow 0, k_1\rightarrow 1$ we get back the metric for pure states calculated above. 

As a second remark let us observe that, in comparing  our results with  those contained in \cite{Ercolessi, Ercolessi2}  different regularization procedures have been employed, which yield different results.  While in \cite{Ercolessi, Ercolessi2}  the starting point is the quantum Fisher information tensor and the symmetric  logarithmic derivative is  employed, we have used the relative Tsallis entropy as a generating function for the metric, which essentially amounts to define the logarithmic derivative through the  introduction of q-logarithms. When the   symmetric logarithmic derivative  is used to compute the quantum Fisher tensor, the calculation involves the trace over three $\mathfrak{u}(N)$  generators (including the identity), thus yielding   symmetric and   antisymmetric terms,  whereas  our definition   in Eq.\eqn{genmet}, consistently with it being directly the metric tensor,  only involves  the trace of two  Lie algebra elements, which can only be symmetric. However, our result has the same  the structure as the symmetric part of the Fisher tensor obtained in \cite{Ercolessi2}, Eq. (55),  in the sense that it splits into three $SU(2)$ related contributions.  

\section{Tomographic metric for qudits}\label{tomo}
In this section we derive the tomographic metric for density matrices associated to $d$-levels quantum systems, the so-called  qudit states. In order to be coherent with the notation used up to now, we shall trade $d$ for $N$, and talk about $N$ level quantum states (because of the action of $SU(N)$) although keeping the shorthand qudits.  In view of this, we shortly review the spin tomographic setting \cite{DodPLA, OMJETP, MarPhSc00, Ventrrev}. 

 The idea of spin tomography is to describe N-level  quantum states by a family of  fair 
probability distributions. We consider a resolution of the identity, say, $\sum_j |e_j\rangle\langle e_j|=\mathbf{1} $, and define $P_j(\rho)= \langle e_j|\rho| e_j\rangle$. We obtain 
in this manner a probability vector associated with every state and every resolution of the identity.
We get a family of probability vectors by using alternative resolutions  which may be obtained from the starting one by means of the action of a unitary operator,  say $\sum_j u |e_j\rangle\langle e_j| u^\dag =\mathbf{1} $, with $u\in SU(N)$. We can compactly write    $\mathcal{W}(m| u), \, -j\le m\le j$ representing the spin projection along  the ``$z$ axis'', $j(j+1)$ being  the eigenvalues of the square of the  angular momentum. 
\be
I= \sum_m |m\rangle\langle m|
\ee
is a  decomposition of the identity as well as 
\be
I_u=\sum_m u  |m\rangle\langle m|u^\dag=  \sum_m |m, u\rangle\langle m, u|
\ee
for each $ u\in SU(N)$. The latter provides a basis in the unitarily rotated reference frame in the Hilbert space, in correspondence of each $u$. Namely, the unitary matrix $u$ labels the reference frame where the spin states are considered. To each state represented by the density matrix $\rho$,  we associate the tomographic probability distribution $\mathcal{W}(m|u)$, through a specific dequantizer operator,  according to the general dequantization  procedure \cite{MMV1, MMV2}  $\rho\rightarrow \mathcal{W}= \Tr \rho D$,  with $D(m,u)=u^\dag |m\rangle\langle m| u\ $ the spin-tomographic dequantizer. The spin-tomographic probability distribution is thus    given by the diagonal matrix elements of the density operator $\rho$, calculated in the given reference frame, that is 
\be
\mathcal{W}(m|u)= \Tr \left( \rho u^\dag |m\rangle\langle m| u\right) = \langle m|u\rho u^\dag | m\rangle . \label{tomogram}
\ee
This can also be interpreted as the conditional probability distribution of the spin projection $m$, provided the unitary matrix $u$, namely  the reference frame, is fixed. Such interpretation  was discussed in \cite{margarita_joint}. 
 
Eq. \eqn{tomogram} is invertible, as shown in \cite{OMJETP,MarPhSc00,FilippovJRusLasRes, Wei1,Wei2}, if a sufficient number of reference frames, so called {\it quorum}, is provided. Each reference frame is determined by choosing a specific group element $u_k$. Given the number of parameters $N^2-1$ which identify the quantum state, a minimal number of reference frames is equal to $N+1$. This can be understood in the following way: $N^2-1$ is the dimension of the Lie algebra of the group $SU(N)$ associated to the $N$-level system. Its Cartan subalgebra, which is associated to the diagonal form of density states, is $N-1$ dimensional. The dimension of a possible  quorum is the number of inequivalent ways one can embed the Cartan subalgebra into the Lie algebra of $SU(N)$. This number is equal to $N+1$. 

\subsection{Tomograms for qubits} 
Let us consider the density matrix associated to a two-level quantum system, as in \eqn{rho2}
\be \label{rrho2}
\rho=\left(
\begin{array}{cc}
\frac{1+y_3}{2} &  \frac{y_1-i y_2}{2}  \\
  \frac{y_1+i y_2}{2} &   \frac{1- y_3}{2}
\end{array}
\right) 
\ee
with $y_1^2+ y_2^2+ y_3^2=   w^2 \le 1$. 
  It is straightforward to check that in the latter case the quorum is equal to  $N+1= 3$ and it is represented by the three rotated reference frames obtained through the unitary matrices
\be
u_{1}= \exp(i \frac{\pi}{4}\sigma_2), \;\;\;  u_{2}= \exp(-i \frac{\pi}{4}\sigma_1) , \;\;	\; u_{3}= {\bf 1}
\ee
which generate, through adjoint action,  rotations of the original reference frame with the spin parallel to the  z axis,    to the reference frames where the spin is parallel to the  $x$, $y$ and (again) $z$ axes respectively.  Indeed we obtain that the probability of measuring the spin projection $m= 1/2$ in each reference frame is given by
\beqa
\mathcal{W}(\frac{1}{2}|u_{1})&\equiv& \langle \frac{1}{2}|u_{1} \rho u_{1}^\dag|   \frac{1}{2} \rangle = \frac{1+y_1}{2}\\
\mathcal{W}(\frac{1}{2}|u_{2})&\equiv& \langle \frac{1}{2}|u_{2} \rho u_{2}^\dag|   \frac{1}{2} \rangle = \frac{1+y_2}{2}\\
\mathcal{W}(\frac{1}{2}|u_{3})&\equiv& \langle \frac{1}{2}|u_{3} \rho u_{3}^\dag|   \frac{1}{2} \rangle = \frac{1+y_3}{2}
\eeqa
and analogously, the probability of measuring $m= -1/2$, $\mathcal{W}(-\frac{1}{2}|u) = 1- \mathcal{W}(\frac{1}{2}|u)$, 
\beqa
\mathcal{W}(-\frac{1}{2}|u_{1})&\equiv& \langle -\frac{1}{2}|u_{1} \rho u_{1}^\dag| -  \frac{1}{2} \rangle = \frac{1-y_1}{2}\\
\mathcal{W}(-\frac{1}{2}|u_{2})&\equiv& \langle -\frac{1}{2}|u_{2} \rho u_{2}^\dag| -  \frac{1}{2} \rangle = \frac{1-y_2}{2}\\
\mathcal{W}(-\frac{1}{2}|u_{3})&\equiv& \langle- \frac{1}{2}|u_{3} \rho u_{3}^\dag| -  \frac{1}{2} \rangle = \frac{1-y_3}{2}
\eeqa
which imply 
\be
y_k = \mathcal{W}(\frac{1}{2}|u_{k})-\mathcal{W}(-\frac{1}{2}|u_{k}) = 2 \mathcal{W}(\frac{1}{2}|u_{k})-1 . \label{xW}
\ee
Thus, the inverse formula for the qubit density matrix is straightforwardly obtained by substituting in Eq. \eqn{rrho2} the above relations,
\be
\rho(\mathcal{W})= \frac{1}{2}\left(\sigma_0 + (2\mathcal{W}_k -1) \sigma_k\right) \label{inversetomogram}
\ee
where we have introduced the notation $ \mathcal{W}_k\equiv \mathcal{W}(\frac{1}{2} |u_k)$.  This means that it is possible to write the density matrices of arbitrary qubit states in terms of the tomographic probabilities $\mathcal{W}(m|u_k)$, which are specified  by the corresponding {\it quorum} of reference frames.

To summarize, Eqs. \eqn{tomogram} and \eqn{inversetomogram} yield, for a generic, mixed, two-level quantum state, the tomographic representation and its inverse formula. From a geometric point of view, we have associated through an invertible map to each point in the Bloch sphere representing the manifold of  quantum states,   a point in the Cartesian product, $K$,  of three 1-simplices of tomographic probabilities. 
To pure states on the boundary of the Bloch sphere, characterized by $y_1^2+ y_2^2+ y_3^3= 1$,  are associated points on   the boundary of the 1-simplex of tomographic probabilities.  For points  on the boundary,  only two reference frames are needed in order to reconstruct the state, for example $u_1$ and $u_3$, so that the probability $\mathcal{W}_2$ is determined by the other two
 \be
 (2\mathcal{W}_2-1)^2= 1- (2\mathcal{W}_1-1)^2 - (2\mathcal{W}_3-1)^2.
 \ee
To complete the geometric picture, we look for a reconstruction formula relating the tomographic Fisher-Rao metric, which is available on the space of tomographic probabilities, to the family of quantum metrics which we have derived in the previous section on the manifold of quantum states. This will be the subject of next section. 

\subsection{Tomograms for qutrits}
Let us consider the density matrix associated to a three-level quantum system, as in \eqn{rho3}
\be \label{rrho3}
\rho= x^j \lambda_j =\left(
\begin{array}{ccc}
y^1&y^2&y^3 \\
y^4&y^5& y^6\\
y^7 &y^8&y^9  
\end{array}
\right) 
\ee
which amounts to 
\beqa
&& y^1=x^0+x^3 +\frac{x^8}{\sqrt 3},\;\; y^2= (y^4)^*= x^1-i x^2 , \;\; y^3= (y^7)^*= x^4-i x^5  \nn\\
&&  y^5=  x^0- x^3 +\frac{x^8}{\sqrt 3},\;\; y^6= (y^8)^*=  x^6-i x^7, \;\; 
 y^9=  x^0-\frac{2x^8}{\sqrt 3} 
\eeqa
with $y^1+y^5+y^9=1$.  Alternatively one can use the Weyl basis for Hermitian matrices, 
\beqa
&& \mathcal{E}^s_{jk}= E_{jk}+ E_{kj}, \;\;  \mathcal{E}^a_{jk}=i ( E_{jk}- E_{kj}),  \;\;  j< k \nn\\
&&  \mathcal{E}^s_{jj}= E_{jj}
\eeqa
where $(E_{jk})_{rs}= \delta_{jr} \delta_{ks}$, 
which is more suitable for generalizations to any dimension.  

  It is straightforward to check that in the latter case the quorum, that is the number of independent bases, is equal to   $N+1= 4$. 

The tomographic probability distribution $\mathcal{W}(m|u)$, where $u$ is a given unitary matrix determining the reference frame, is given by
\be
 \mathcal{W}(m|u)= \langle m|u\rho u^\dag|m\rangle\;\; m=1,0,-1 
 \ee
 The spin-one projection $m$ on the ``z axis'' has the probability $\mathcal{W}(1|u=I)= y_1$, $\mathcal{W}(0|u=I)= y_5$, $\mathcal{W}(-1|u=I)= y_9$. The tomogram $\mathcal{W}(m|u)$ represents the probability of having projection $m$ in the rotated reference frame. It can be expressed 	by using the matrix-vector map as  \cite{sudama}   
 \beqa \label{sis}
 \mathcal{W}(1|u)&=& \sum_{k=1}^9 \mathcal{U}_{1k} y^k \nn\\
 \mathcal{W}(0|u)&=& \sum_{k=1}^9 \mathcal{U}_{5k} y^k \nn\\
 \mathcal{W}(-1|u)&=& \sum_{k=1}^9 \mathcal{U}_{9k} y^k 
  \eeqa
  where the unitary $9\times 9$ matrix $\mathcal{U}= u\otimes u^*$ and $u^*$ is the complex conjugate matrix of $u$.  In order to express the density matrix $\rho$ in terms of tomographic probabilities we may use four matrices $\mathcal{U}^{(k)}= u_k\otimes u^*_k, k=1,..,4$, which give a  sufficient set. The $3\times 3$ matrices $u_k$ can be chosen as matrices $u_k(\phi,\theta,\psi)=u(\phi_k,\theta_k,\psi_k) $ of an irreducible spin one representation of $SU(2)$. As a result, the matrices $u_k$ depend only on two of the three Euler angles $(\phi_k,\theta_k,\psi_k)$.      The $u_k$ are determined by four unit vectors $n_k\equiv(\sin\theta_k \cos \phi_k, \sin\theta_k \sin \phi_k , \cos\theta_k)$ so that 
\be
  \mathcal{W}(m|u_k)=   \mathcal{W}(m|n_k)
  \ee
  Since $\sum_m \mathcal{W} (m|n_k)=1$ we can choose $\mathcal{W} (1|n_k)$ and $\mathcal{W} (0|n_k)$ as the independent ones, so as to have 8 independent probabilities obtained on varying $n_k$.
  We pose 
  \be
 \mathcal{W}(1|n_k)= \mathcal{W}_{2k-1}, \;\; \mathcal{W}(0|n_k)= \mathcal{W}_{2k},
 \ee
  with $k=1,...,4$.
 In view of Eqs. \eqn{sis} we may write the equations connecting the eight independent  parameters of the density matrix ($y^9=1-y^1-y^5$) as
 \beqa
   \mathcal{W}_{2k-1}&=& {\sum_{1}}^8 \mathcal{U}^{(k)}_{1j} y^j+ \mathcal{U}^{(k)}_{19}(1-y^1-y^5) \\
 \mathcal{W}_{2k}&=& \sum_{1}^8 \mathcal{U}^{(k)}_{5j} y^j+ \mathcal{U}^{(k)}_{59}(1-y^1-y^5)
 \eeqa
  with $k=1,...,4$. 
  By posing $\vec{ \mathcal{W}}= ( \mathcal{W}_1,..., \mathcal{W}_8)^T$ and   
  \be
  \vec y_{red}= (y^1,...,y^8)^T
  \ee 
  we have 
  \be
  \vec{ \mathcal{W}}= A \, \vec y_{red}+ \vec B
  \ee
  where
  \be
  B_{2k-1}= \mathcal{U}^{(k)}_{19}; \;\; B_{2k}= \mathcal{U}^{(k)}_{59} 
  \ee
  while the rows of the matrix $A$ read 
  \be
  A_{2k-1,j}= \mathcal{U}^{(k)}_{1j}-\mathcal{U}^{(k)}_{19} (\delta_{j1}+\delta_{j5}); \;\;  A_{2k,j}= \mathcal{U}^{(k)}_{5j}-\mathcal{U}^{(k)}_{59} (\delta_{j1}+\delta_{j5})
  \ee
  with $k=1,...,4$. 
  If $A$ is non singular we recover the parameters of the density matrix $y^j$ as a linear combination of the tomograms. We recall that $ \mathcal{U}^{(k)}=u_k\otimes u_k^*$. For particular choices of the unitary matrices $u_k$ the matrix $A$ can have diagonal form. This corresponds to a choice of four different realizations of the Cartan subalgebra of $SU(3)$ where the basis generators are mutually orthogonal to each other. This amounts to single out four unit vectors  $n_k, k=1,...,4$ such that, measuring the probabilities of spin projections $m=1,0$ on these directions, provides  the expression of the elements $y^j(\vec{ \mathcal{W}})$ of the density matrix. 
  
\subsection{The tomographic metric and the reconstruction formula}
In this section we address the following problem. The qudit density matrix is uniquely reconstructed if  the tomographic probabilities of the qudit are known. On the other hand  the geometry of the tomographic probability space  is  known and the Fisher-Rao metric is uniquely singled out by the monotonicity property (see \cite{fujiwara} for a recent discussion in the qubit case). Is it possible to reconstruct the quantum metric calculated in the previous section in terms of the ``classical'' tomographic Fisher-Rao metric? 

In order to understand such issues in a simple setting,    we shall concentrate  on  the two-dimensional case. 
The question above is related to another one. The quantum states contain information on phases of the complex wave vector.  Probability distributions such as the tomograms of qubit states have no phases, but they determine the quantum states completely.  What is the mechanism by which we recover the phases from nonnegative probability distributions? We can understand this mechanism looking at Eq. \eqn{inversetomogram}. The off-diagonal elements of the density matrix encode the phases both for mixed and pure states. That is, phases are determined by probabilities of spin projections when measured in all  reference frames of a chosen sufficient set. Therefore, the role of the chosen
 family of reference frames is crucial and specifies how to recover the quantum state from classical tomographic probabilities.  

In order to compute the tomographic metric we follow the approach described in the previous sections.  Let us consider the relative Tsallis entropy associated to tomographic probabilities
\be
0\le S_q(\rho, \tilde{\rho}, u, \tilde u)= \left(q(1-q)\right)^{-1}\left(1-  \sum_m {\mathcal{W}_\rho}^q(m |u) \,  {\widetilde{\mathcal{W}}_{\tilde{\rho}}}^{1-q}(m |\tilde u)\right)
\ee
with $m=\pm1/2$, $\rho, \tilde{\rho}$ corresponding to two different states, with their reference frames, $u, \tilde u$. The metric is obtained from an analogous expression to \eqn{genmet} 
\be
G=-i^*d\, \tilde d S_{q}(\mathcal{W}_\rho, {\widetilde{\mathcal{W}}_{\tilde{\rho}}})=\left(q(1-q)\right)^{-1}\sum_m d  {\mathcal{W}_\rho}^q \otimes d  {\mathcal{W}_{\rho}}^{1-q}=  \sum_m \mathcal{W}_\rho d\ln\mathcal{W}_\rho \otimes d\ln\mathcal{W}_\rho 
\ee
which, as expected,  has the form of the Fisher Rao metric on the space of tomographic probabilities. 
This metric can be expressed in terms of the parameters characterizing the density matrix, as follows 
\be
G  (y, u)= G_{jk}(y,u) dy_j\otimes dy_k \label{tomometric}
\ee
with 
\be
G_{jk}  (y, u)= - \frac{\del^2 S_q}{\del y_j\del \tilde y_k} (\rho=\tilde{\rho}, u=\tilde u) = \frac{1}{\mathcal{W}(1/2|u) \mathcal{W}({-1/2|u})} C_{jk} (u) \label{capitalG}
\ee 
where the matrix $C_{jk}$ depends on the reference frame, i.e. only on the unitary matrix $u$. It reads
\be 
C=
\left(
\begin{array}{ccc}
 ( Re (u_{11} u_{12}^*))^2  & Im (u_{11} u_{12}^*)  Re(u_{11} u_{12}^*) & Re(u_{11} u_{12}^*)(|u_{11}|^2 -\frac{1}{2})  \\
 Im (u_{11} u_{12}^*)  Re(u_{11} u_{12}^*) &(Im (u_{11} u_{12}^*))^2&( Im (u_{11} u_{12}^*))(|u_{11}|^2 -\frac{1}{2})   \\
Re(u_{11} u_{12}^*)(|u_{11}|^2 -\frac{1}{2}) &( Im (u_{11} u_{12}^*))(|u_{11}|^2 -\frac{1}{2})& (|u_{11}|^2 -\frac{1}{2})^2
\end{array}
\right)
\ee
Three unitary matrices specifying the quorum of reference frames are
\be
C_{jk}(u_1)= \frac{1}{4}  \delta_{j1}\delta_{k1}, \;\;\; C_{jk}(u_2)= \frac{1}{4}  \delta_{j2}\delta_{k2},\;\;\; 
C_{jk}(u_3)= \frac{1}{4}  \delta_{j3}\delta_{k3}
\ee
with all the other entries equal to zero. On replacing in Eq. \eqn{capitalG} we obtain
\be
G_{jk}  (u_l)= \frac{1}{4}\frac{1}{\mathcal{W}_l(1-\mathcal{W}_l)} \delta_{jk}\delta_{kl}
\label{Gjk}
\ee
from which we get
\be
\mathcal{W}_k=\frac{1\pm \sqrt{1-1/G_{kk}}}{2}
\ee
with $G_{kk}$ the diagonal elements of the matrix \eqn{Gjk}. Thus, each one of the tomographic metrics is  degenerate. However, from these degenerate metrics one can reconstruct the metric obtained from the quantum relative entropy \eqn{n=2q=1metric}
Indeed, recalling that 
\be
2 \mathcal{W}_k-1 = y_k
\ee
we get 
\be
y_k= \pm\sqrt{1-G_{kk}^{-1}}.
\ee
This provides the expression of the quantum metrics $g_q (y)$ in terms of the tomographic metric $G (u)$.

\subsection{Uncertainty relations for tomographic probabilities}\label{uncert}
Since the density matrix of qubit states is expressed in terms of probabilities $\mathcal{W}_j, j=1,2,3$ of spin projections $m=\pm 1/2$ measured in three orthogonal reference frames, the non-negativity condition for the determinant of the matrix \eqn{inversetomogram}   yields the inequality
\be
\left(\mathcal{W}_1-\frac{1}{2}\right)^2+  \left(\mathcal{W}_2-\frac{1}{2}\right)^2 + \left(\mathcal{W}_3-\frac{1}{2}\right)^2 \le \frac{1}{4}. \label{tomouncert}
\ee
The latter  represents the uncertainty relations reflecting the quantum correlations of qubits for spin projections measured in three different directions. 

As discussed above, the qubit density matrix is mapped onto points of $K$, the Cartesian product of  three different 1-simplices which are determined by probabilities $0\le \mathcal{W}_j, j=1,2,3$. However the domain of the probabilities $\mathcal{W}_j$ is not the whole $K$,   the ``cube'' determined by the three intervals. While each one of the parameters $\mathcal{W}_j$  satisfies  $0\le\mathcal{W}_j \le 1$, the set of parameters $\mathcal{W}_j$ corresponding to quantum qubit states is further restricted by the uncertainty relation \eqn{tomouncert}. 

This inequality can be checked experimentally. One can measure spin projections $m=\pm 1/2$ in three orthogonal directions in three-dimensional space. For instance, it takes place  for qubits realized by two-level atoms or by superconducting qubits realized by Josephson junctions \cite{Glush, Fedorov, Martinis, Astaf, Devoret}. The possibility to check experimentally the inequality given by Eq.(\ref{tomouncert})
is connected with the development of devices (superconducting
circuits) based on Josephson junctions. They were used, for example, to check
the dynamical Casimir effect (non-stationary Casimir effect) \cite{DMM89,FMHKZ, OM94}. 
 The
vibrations of current and voltage in the superconducting circuits are
described by the model of a quantum oscillator. Experimentally only the first few
levels of the oscillator are excited. Due to this the states of the system 
practically are qubit or qudit states. Such qubits in
superconducting circuits can be obtained in the mixed states with the
density matrix  given by Eq. (\ref{inversetomogram}). Three probabilities determining the
density matrix must satisfy the uncertainty relation formulated as
inequality given by Eq. (\ref{tomouncert}). This fact reflects the presence of quantum
correlations in the one-qubit state. In the case of two qubits in entangled
states the quantum correlations are responsible for the violation of Bell
inequalities. The measuring the three probabilities determining one qubit
state density matrix provides the possibility to confirm the presence of
quantum correlations in superconducting circuits for one-qubit state.
Analogously, for a superconducting qutrit the measuring the
tomographic probabilities for spin 1 projections, determining the density
matrix of the system, can check the presence of quantum correlations
expressed in terms of a non-negativity condition  of the eigenvalues of the
density matrix depending on these probabilities.

 Analogous inequalities can be obtained for qudits.  For instance in the three dimensional case the  inequalities stemming out from the non-negativity of the $3\times 3$ density matrix read $y_1(\vec{\mathcal W}) y_2(\vec{\mathcal W})-y_4(\vec{\mathcal W}) y_5(\vec{\mathcal W})\ge 0$ and that $\rho({\bf y}_{\vec{\mathcal W}})\ge 0$.  All these relations can be checked experimentally. 
 
 Points with coordinates $\mathcal{W}_j, j=1,2,3$, satisfying the inequality \eqn{tomouncert} belong to the three-dimensional ball of radius $1/2$ and center $(1/2, 1/2,1/2)$. The ball is inscribed into the surface of the cubic  complex of edge equal to $1$.  Points belonging to the boundary of the ball  saturate the inequality and correspond to pure states  whereas points inside represent mixed states. Notice that  the tomographic metric \eqn{tomometric} does not coincide with the ``natural metric'' in the ball, which would be the Euclidean metric. Indeed, the latter is related to the distance function
 \be
 s=\sqrt{(\mathcal{W}_1-\widetilde{\mathcal{W}}_1)^2+(\mathcal{W}_2-\widetilde{\mathcal{W}}_2)^2+(\mathcal{W}_3-\widetilde{\mathcal{W}}_3)^2}
 \ee
which, however, does not fulfill the monotonicity request. The monotonicity  request is justified by  the physical demand that, in processes of dissipation associated to the evolution of a system induced by interaction with the environment, the distinguishability  of  quantum states can not increase \cite{Holevo}. This means that  completely positive maps which describe such process  can only decrease the distance. As it is well known,  the monotonicity property gives rise to the  uniqueness of the Fisher-Rao metric as shown by  Chentsov \cite{Chentsov82}. 

Distinguishability of two quantum states $\rho$ and ${\tilde{\rho}}$ may be described by the two-points function  known as fidelity
\be
\Phi = \Tr \sqrt{\rho  \tilde{\rho}} .
\ee
 In general the fidelity is used to describe  a kind of distance between two states, namely, two states are close to each other if their fidelity is close to 1 \cite{Chuang}, however some critical remarks about such interpretation can be found in \cite{Dod, Paris}. 
Here we observe that the function $\Phi$,   although symmetric,  should be properly related to a divergence function rather than to a metric.  Indeed  a  monotonic potential function is represented by
\be
\mathfrak{S}_{\frac{1}{2},\frac{1}{2}} (\rho, \tilde{\rho}) =  (1-\Tr \sqrt{ \rho \tilde{\rho}} )
\ee
namely the   quantum analogue of   the function $F$ in Eq. \eqn{altpot}. This is  an instance of the so called $\alpha$-$z$ entropies \cite{datta}, $\mathfrak{S}_{\alpha,z}$, with $\alpha=z=1/2$, which are a generalization of relative Tsallis entropy.  Let us notice that the related metric is the  Wigner-Yanase metric discussed for instance in  \cite{Gib1, Gib2}. We shall consider this large family of   divergence functions  in a forthcoming paper.

The different metrics which one can introduce to describe the distinguishability of quantum states, e. g. of qubits, can reflect different choices of tomographic descriptions of quantum states. Once a choice has been made, the Fisher-Rao metric on ``fair probability distributions'' is uniquely determined, therefore it may appear puzzling  that, by combining these metrics, one may obtain an infinite family of alternative ``quantum metrics''.  To clarify how this could happen, let us consider a simple analogy.  Given  $(x_1, x_2)$, Cartesian coordinates for points on the plane, we can introduce alternative partial metrics with two different, degenerate tensors
\be
g^{(1)}= \left(\begin{array}{cc}
1&0\\0&0\end{array}\right); \;\;\; 
g^{(2)}= \left(\begin{array}{cc}
0&0\\0&1\end{array}\right)
\ee
Thus, the distances between two points $(x_1,x_2), (\tilde x_1,\tilde x_2)$ on the plane are given by the partial metrics as
\be
ds_1^2=d x_1^2; \;\;\; ds_2^2= dx_2^2
\ee
amounting to define the distances in terms of the projection on the abscissa or ordinate axes. In this metrics different vectors with the same projection on one of the axes  will be indistinguishable. When we try to  combine the two partial metrics  it is possible to use different coefficients, say, $ds^2= f_1^2 dx_1^2+ f_2^2 d x_2^2$, depending on the projection from $\R^2$ to $\R$, so as to obtain a large class of metrics among which  the whole non degenerate tensor $g_{ij}= \delta_{ij}$. 

\section{Gaussian symplectic tomograms}\label{sympl}
In this section we discuss  Gaussian probability distributions to better distinguish the role of statistical manifolds with respect to statistical models. Here, the statistical manifold is infinite-dimensional while a statistical model,  the finite dimensional manifold  of Gaussian distributions, is embedded into the infinite dimensional statistical manifold of symplectic tomograms. The symplectic tomogram of a density state $\rho$   \cite{Ventrrev,introtom} 
\be
{\mathcal W}_\rho (X,\mu,\nu)= \Tr\left[ \hat \rho\, \delta(X \textbf{1}-\mu \hat q -\nu \hat p)\right]
\ee
where $\hat q, \hat p$ are usual position and momentum operators, X is the random variable while $\mu,\nu$ are real parameters. The inverse formula is  
\be
\hat \rho= \int \frac{\dd\mu\;\dd\nu}{2\pi}\; dX\,  {\mathcal W}_\rho (X,\mu,\nu) e^{i (X \textbf{1}-\mu \hat q -\nu \hat p) }.
\ee
The matrix elements of the density matrix in the Fock basis $|n\rangle$ are functionals of the tomographic probability distribution ${\mathcal W}_\rho (X,\mu,\nu)$,
\be
\rho_{nm}= \int \frac{\dd\mu\;\dd\nu}{2\pi}\; dX\, e^{iX } {\mathcal W}_\rho (X,\mu,\nu) \langle n| e^{-i (\mu \hat q +\nu \hat p) }| m\rangle .  \label{rhonm}
\ee
The matrix elements in Eq. \eqn{rhonm} can be expressed in terms of Laguerre polynomials \cite{glau}. 

The tomogram of a Gaussian state of a one-mode photon field is 
\be
{\mathcal W}_\rho (X,\mu,\nu)= \frac{1}{\sqrt {2\pi \sigma}} e^{-\frac{(X-\bar X)^2}{2\sigma}}
\ee
The parameters $\sigma$ and $\bar X$ are respectively the dispersion and the mean value of the homodyne quadrature $X$ in the reference frame determined by $\mu,\nu$, namely
\beqa
\bar X&=& \mu \langle \hat q \rangle + \nu \langle \hat p \rangle \label{Xbar}\\
\sigma&=& \mu^2 \sigma_{qq} + \nu^2 \sigma_{pp} + 2 \mu\nu \sigma _{qp} \label{sigma}
\eeqa
where
\beqa
\sigma_{qq}&=& \Tr\rho \hat q^2 -(\Tr \hat \rho \hat q)^2:= y^1 \ge 0 \\
\sigma_{pp}&=& \Tr\rho \hat p^2 -(\Tr \hat \rho \hat p)^2:= y^2 \ge 0 \\
\sigma_{qp}&=& \Tr\rho \frac{\hat q\hat p+ \hat p \hat q}{2} -\Tr \hat \rho \hat q \Tr \hat \rho \hat p := y^3 \\
\langle \hat q \rangle &=& \Tr \hat \rho \hat q := y^4 \;\;\;  \langle \hat p \rangle = \Tr \hat \rho \hat p := y^5
\eeqa
span the space of parameters. Thus,  for a Gaussian state a density matrix is determined by five parameters  which, for classical Gaussian states, satisfy 
\be
\sigma_{qq} \sigma_{pp}- \sigma_{qp}^2 \ge 0 \label{cline}
\ee
while for quantum states 
\be
\sigma_{qq} \sigma_{pp}- \sigma_{qp}^2 \ge \frac{1}{4}, \label{quine}
\ee
which is the Schr\"odinger-Robertson uncertainty relation \cite{Rob,Sch}.  We recover the analogue of the inequality previously discussed for qubits \eqn{tomouncert}.
The metric for Gaussian density states has been discussed in \cite{mancinipol}.  In the tomographic setting, the Tsallis relative entropy for two Gaussian states with parameters ${\bf y}=
\{y^j\} , \tilde{\bf y}=\{\tilde y^j\}$ is
\be
S_q({\bf y},  \tilde{\bf y}, \mu, \nu)= \left(q(1-q)\right)^{-1} \left[1-\int \dd X\, {\mathcal W}^q(X, \mu, \nu;  {\bf y}) {\mathcal W}^{1-q}(X, \mu, \nu, \tilde {\bf y}) \right].
\ee
In view of the Gaussian form of the tomograms the Tsallis entropy is readily evaluated by completing the square as
\be
\int \dd X \frac{1}{\sqrt{2\pi \sigma^{q}  \tilde \sigma^{1-q}}} e^{-A X^2+ B X } =  \frac{1}{\sqrt{2\pi \sigma^{q}  \tilde \sigma^{1-q}}} \sqrt{\frac{\pi}{A}} e^{\frac{B^2}{4 A}}
\ee
where 
\be
A=  \frac{q}{2\sigma}+ \frac{1-q}{2\tilde\sigma}, \;\; \; B=  \frac{q\bar X}{\sigma}- \frac{(1-q)\bar {\tilde X}}{\tilde\sigma}
\ee
and $\sigma, \bar X , \tilde \sigma, \tilde X$ are given in Eqs. \eqn{Xbar}, \eqn{sigma} in terms of the parameters $y^j, \tilde y^j$ respectively. The explicit form of the metric is thus easily obtained by a lengthy but straightforward calculation by differentiating the Tsallis entropy with respect to the parameters $y^j, \tilde y^j$:
\be
G^{symp}_{jk}({\bf y, \mu, \nu}) = -\frac{\del^2}{\del y^j \del \tilde y^k} S_q ({\bf y}, \tilde {\bf  y}, \mu,\nu) \vert_{{\bf y}= \tilde {\bf y}}.
\ee
The Fisher-Rao metric for  symplectic tomograms is obtained in the limit $q\rightarrow 1$ and it reads
\be
\delta S^2= \int \dd X \, {\mathcal W}(X,\mu,\nu) \delta \ln {\mathcal W}(X,\mu,\nu) \otimes \delta\ln {\mathcal W}(X,\mu,\nu). 
\ee
Let us notice that the classical tomographic metric is different from the quantum one, even though they can coincide locally, because the statistical models parametrized in terms of $y^j$'s  are different in view of the different inequalities Eqs. \eqn{cline}, \eqn{quine} satisfied by them.

\section{Conclusions}
To summarize, we point out the main results of our work. Starting form the relative Tsallis $q$-entropy as a potential function for the metric, we derived a one-parameter family of quantum metrics for full rank $N$-level systems, Eq \eqn{generalg} and analyzed in detail the cases $N=2,3$, Eqs. \eqn{k1k2met}, \eqn{gtgorb}.  Metrics for lower rank strata of states (such as pure states) have been obtained by means of the radial limit procedure. Specific  limit values of q have been studied and some known metrics, such as the Wigner-Yanase metric,   have been recovered.   Let us stress that, in order to obtain these results we had to define a differential calculus adapted to the non-commutative case, which can be extended to other families of potential functions. 

Then we constructed explicitly the Fisher-Rao tomographic metric for qubit and qutrit states in  different reference frames on the Hilbert space of quantum states.    We thus  expressed the quantum metric derived from  the quantum Tsallis entropy  for the $N=2$  case in terms of parameters which are tomographic probabilities of spin 1/2 projections onto three perpendicular directions in the space $\R^3$. To our knowledge such invertible maps between tomographic parameters and parameters identifying density states are here discussed for the first time.  Starting from the usual formulation of uncertainty relations, we have  obtained new constraints  for quantum  tomograms. The new relation could be checked experimentally, either for spin 1/2 particles or for two-level atoms, or for superconducting qubits realized in experiments with Josephson junction devices.   
The comparison of different metrics associated with different kinds of tomographic schemes will be considered in a future publication. 

The behaviour of the connection and the curvature, as computed in the appendices, shows that the non-vanishing of the curvature is essentially due to the quantum contribution, i.e., the part along the orbits of the unitary group. The negativity of the scalar curvature means that autoparallel curves do not cross more than once,  and therefore any two states may be connected by an unique autoparallel curve.

\noindent{\bf Acknowledgements} 

The authors are deeply indebted with Marco Laudato and Fabio Mele for their critical reading of the manuscript and useful discussions, which greatly improved the paper. 
G.M.  would like to acknowledge the grant ``Santander-UC3M Excellence Chairs 2016Ó. P.V.  acknowledges  support by COST (European Cooperation in Science  and  Technology)  in  the  framework  of  COST  Action  MP1405  QSPACE.
V. I. M. thanks the Physics Department at Naples U. and the local INFN section  for hospitality and financial support. 

\setcounter{section}{0}
\appendix
\section{Dual connections}\label{appa}
In order to compute the connections it is convenient to rewrite the relative entropy function using polar coordinates $\theta, \phi$ on the spherical orbits, together with the parameter $w$ for the transversal part. 
Since $\rho^q= U\rho_0^q U^{-1}$, on using Eqs. \eqn{rho0q}, \eqn{rho01-q} and \eqn{xes} we observe that 
\beqa
\rho^q&=& \frac{1}{2}\left[ (a_q+b_q) \sigma_0+ (a_q-b_q)   \bf{x}\cdot\bf{\sigma} \right]\\
\tilde \rho^{1-q}&=& \frac{1}{2}\left[ (\tilde a_{1-q}+\tilde b_{1-q}) \sigma_0+ (\tilde a_{1-q}-\tilde b_{1-q})  \bf{\tilde x}\cdot \mathbb{\sigma} \right]\,.
\eeqa
We have then 
\be
S_{Ts} (\rho, \tilde{\rho})= \left(q(1-q)\right)^{-1} \left\{1- \frac{1}{2} (a_q+ b_q) (\tilde a_{1-q}+\tilde b_{1-q}) - \frac{1}{2} (a_q- b_q) (\tilde a_{1-q}-\tilde b_{1-q}) \bf{x}\cdot\bf{\tilde x} \right\}
\ee
with $ \bf{x}\cdot\bf{\tilde x}= \sin \theta\sin\tilde\theta\cos(\phi- \tilde\phi) + \cos\theta\cos\tilde\theta$. 

Let us compute the connection coefficients $\Gamma_{jkl}, \Gamma^*_{jkl}$ with respect to the coordinates $w, \theta,\phi$, which will be labelled as $1,2,3$ in the following.
As for the coefficients involving only polar coordinates, they yield an auto-dual connection. We have indeed that,  for $j,k,l\in\{ \theta,\phi\}$,  the only non-zero coefficients  read
\beqa
\Gamma_{233}&=& \Gamma^*_{233} = \frac{1}{4\left(q(1-q)\right)}\left[  \left(\frac{1-w}{1+w}\right)^{q}(1+w)+\left(\frac{1+w}{1-w}\right)^{q} (1-w)-2 \right] {\cos\theta\sin\theta}\nn\\
\Gamma_{323}&=&\Gamma^*_{332}= \frac{1}{4\left(q(1-q)\right)}\left[2-   \left(\frac{1-w}{1+w}\right)^{q}(1+w)-\left(\frac{1+w}{1-w}\right)^{q} (1-w) \right] {\cos\theta\sin\theta}.
\eeqa
As for the others, all coefficients involving a single derivative with respect to the angular coordinates  are zero. Therefore we are left with 
\beqa
\Gamma_{111}&=& 2  (1-q)\frac{w}{(1-w^2)^2} \nn\\
\Gamma^*_{111}&=& 2 q  \frac{w}{(1-w^2)^2}\nn\\
\Gamma_{122}&=& \frac{1}{4 q}\left[ \left(\frac{1 - w}{1+w}\right)^{q} -  \left(\frac{1 + w}{1-w}\right)^{q} \right]
\nn\\
 \Gamma^*_{122}&=& \frac{1}{4(1-q)}\left[ \left(\frac{1 + w}{1-w}\right)^{1-q} -  \left(\frac{1 - w}{1+w}\right)^{1-q} \right]\nn\\
\Gamma_{133}&=& {\sin^2\theta} \, \Gamma_{122}    \nn\\ 
\Gamma^*_{133}&=& {\sin^2\theta} \,\Gamma^*_{122}\nn\\
\Gamma_{212}&=&\frac{1}{4(1-q)} \left[\left(\frac{1+w}{1-w}\right)^{1-q}-\left(\frac{1-w}{1+w}\right)^{1-q}\right]\nn\\
\Gamma^*_{212 }&=& \frac{1}{4 q}\left[ \left(\frac{1 + w}{1-w}\right)^{q} -  \left(\frac{1 - w}{1+w}\right)^{q} \right]\nn\\
\Gamma_{313}&=&{\sin^2\theta}\, \Gamma_{212}\nn\\
\Gamma^*_{313 }&=&{\sin^2\theta}\,\Gamma^*_{212 }\nn\\
 \eeqa
and they are symmetric in the exchange of the last two indices.
The skewness tensor $T_{ijk} $ is therefore different from zero.  Let us notice that the connection is autodual for $q=\frac{1}{2}$, namely $\alpha=0$ in the context of $\alpha$ divergence functions, in agreement with \cite{Am00}. 
The value $q=\frac{1}{2}$ yields  symmetric coefficients, namely, the Levi-Civita connection.

\section{Curvature tensor}\label{appb}
Let us compute the curvature tensor for the connection just calculated. We use   the metric given in Eq. \eqn{n=2metric} and its inverse to lower and raise indices. 
It is defined as 
\be
R^k_{lmn}= \del_m \Gamma^k_{nl}-\del_n \Gamma^k_{ml}+ \Gamma^e_{nl}\Gamma^k_{me}-\Gamma^e_{ml}\Gamma^k_{ne}
\ee
Only fourtheen  of them are different of zero, of which, only seven are independent, the Riemann tensor being antisymmetric in the exchange of the last two indices. They are
\beqa
R^1_{212}&=&\frac{1}{4q} \left\{2 q + \left(\frac{1+w}{1- w}\right)^q \left[\left(q-2\right) w +q\right]- 
 \left(\frac{1-w}{1+ w}\right)^q \left[\left(q-2\right) w -q\right]\right\} \nn\\
 R^1_{313}&=&   \frac{ \sin ^2   \theta}{4 }  \left [2+ \left(\frac{1+w}{1-w}\right)^q \left(1+\left(1-\frac{2}{q}\right)w\right)  +   \left(\frac{1-w}{1+w}\right)^q\left(1-\left(1-\frac{2}{q}\right)w\right) \right]\nn\\
 R^1_{323}&=&\frac{\sin 2 \theta}{4q} \left[\left(\frac{1+w}{1-w}\right)^q (1-w )- \left(\frac{1-w}{1+w}\right)^q (1+w )\right] \nn\\ 
    R^2_{112}&=& \frac{q \left(1-\frac{2 (1- w )^{q-1}}{(1- w )^q-( w +1)^q}\right) \left(-\frac{2
   q (1- w )^{q-1}}{(1- w )^q-( w +1)^q}-\frac{2 (q-1)
    w }{ w -1}+q\right)}{( w +1)^2}\nn\\
  R^2_{323}&=& \cos^2\theta-  \frac{\sin^2\theta}{4}\left[2 + \left(\frac{1-w}{1+w}\right)^q(1+w) +\left(\frac{1+w}{1-w}\right)^q(1-w)\right]
   \nn\\
   R^3_{113}&=& \frac{q}{(1-w^2)^2} \left[  \left( w + \frac{(1-w)q+ (1+w)^q}{(1-w)^q-(1-w)^q}\right)\right] \left[ q \left( w + \frac{(1-w)q+ (1+w)^q}{(1-w)^q-(1-w)^q}\right)- 2(q-1)w \right]\nn\\
    R^3_{223}&=&\cot^2\theta+  \frac{1}{4}\left[2 + \left(\frac{1-w}{1+w}\right)^q(1+w) +\left(\frac{1+w}{1-w}\right)^q(1-w)\right]
      \eeqa
  In the limit $q\rightarrow 1$ they simplify considerably, reading
 \beqa
R^1_{212}&=&1 \nn\\
 R^1_{313}&=&  \sin^2\theta  \nn\\
 R^1_{323}&=& 2 w \cos\theta\sin\theta  
   \nn\\ 
    R^2_{112}&=& \frac{1}{w}\nn\\
  R^2_{323}&=& \cos^2\theta- \sin^2\theta  \nn\\
   R^3_{113}&=& \frac{1}{w^2} \nn\\
    R^3_{223}&=&1+\cot^2\theta.
      \eeqa  
We obtain for the scalar curvature
 \be
\lim_{q\rightarrow 1}     \mathcal{R}=  2\left(1- \frac{1}{w^2}\right) \le 0, \,\,\, |w|\le 1. \label{Rq1}
   \ee
 We can repeat the calculation for the dual connection. For the scalar curvature we obtain
\be
\lim_{q\rightarrow 1}  \mathcal{R}^*= \frac{4 \tanh ^{-1}w \left[w-\left(1-w^2\right) \tanh
   ^{-1}w\right]-2}{\left(1-w^2\right) \tanh
   ^{-1}w^2}. \label{Rstq1}
 \ee
In the limit $q\rightarrow 0$ the results  \eqn{Rq1} and \eqn{Rstq1} get exchanged namely
\be
   \lim_{q\rightarrow 0}  \mathcal{R}=\frac{4 \tanh ^{-1}w \left[w-\left(1-w^2\right) \tanh
   ^{-1}w\right]-2}{\left(1-w^2\right) \tanh
   ^{-1}w^2} \label{R0q}
 \ee
   and
   \be
    \lim_{q\rightarrow 0}  \mathcal{R}^*=2\left(1- \frac{1}{w^2}\right) \label{Rstq0}.
  \ee  
Finally, in the autodual case corresponding to $q=1/2$  (i.e. the Wigner Yanase metric) we find
   \be
 \lim_{q\rightarrow 1/2}  \mathcal{R}=  \lim_{q\rightarrow 1/2} \mathcal{R}^*=\frac{1}{2}- \frac{ \left(\sqrt{1-w^2}+1\right)}{w^2}.
    \ee
  already discussed in \cite{Gib2}.

\end{document}